\documentclass[preprint,12pt]{elsarticle}

\usepackage{color}
\usepackage{subfigure} 

\usepackage{amssymb}
\usepackage{placeins}

\journal{Physica A}

\begin{document}

\begin{frontmatter}



\title{Entangled Economy: an ecosystems approach to modeling systemic level dynamics}


\author[JDR]{Juan David Robalino\fnref{fn1}} 
\ead{jr872@cornell.edu}
\author[HJJ]{Henrik Jeldtoft Jensen\corref{cor1}}
\ead{h.jensen@imperial.ac.uk}

\cortext[cor1]{Corresponding author} 
\fntext[fn1]{Present address: Cornell University - Department of Economics, Ithaca N.Y. 14853, USA}
 
\address[JDR]{Department of Mathematics, Imperial College London, London, SW7 2AZ, UK}
\address[HJJ]{Department of Mathematics and Complexity \& Networks Group, Imperial College London, London, SW7 2AZ, UK}

\begin{abstract}
We present a model of an economy inspired by individual based model approaches in evolutionary ecology. We demonstrate that evolutionary dynamics in a space of companies interconnected through a correlated interaction matrix produces  time dependencies of the total size of the economy total number of companies, companies age and capital distribution that compares well with statistics for USA. We discuss the relevance of our modeling framework to policy making. 
\end{abstract}


\end{frontmatter}


\section{Introduction}
\label{Introd}
Economies are highly complex systems. History suggests that the standard analysis of an economy with the reductionist approach of individual rationality and utility or profit maximization misses important features about aggregate dynamics and global stability that result from the interactions of economical agents \cite{arthur1999complexity}. Complexity theory offers an interesting alternative that has potential to provide better insights about the systemic risks by analyzing the properties of the economy as a whole \cite{Sornette_2004}. 

In this paper we develop and explore a simple model of an economy inspired by complexity models of evolutionary ecology, in particular the work by Laird and Jensen \cite{lair05:tang}\cite{lair07:EcoMod}. This kind of analysis is relatively new in the economics literature, especially when it comes to policy design. In a world increasingly globalized with greater interdependence of the individual economies, analyzing systemic risks becomes critical in order to control and avoid global crisis. Andrew G Haldane, Executive Director of Financial Stability department of the Bank of England, has published numerous papers about systemic risk of the financial sector in which he discusses interesting and innovative ideas for policy design \cite{Haldane2009}\cite{Haldane2011}. Many of his arguments are based on complexity models of the global banking system where system parameters, such as the density of interbank loans, have a drastic effect on the systemic risk. In general these complexity models consist of units interacting and forming networks. Important tendencies of the systemic risk can be drawn from properties of the network and the dynamics of the system such as the number of links or hierarchy of the network.  Such properties could be taken in to account when designing economic policies and regulations. The first step to this aim is to establish a minimal modeling framework of the economy that allows us to reproduce, at least at a qualitative level, dynamical behavior consistent with real statistics, and that can be used to develop an appropriate concrete ecological approach to economics allowing us to rely less on purely biological analogies \cite{Johnson2011}\cite{Lux2011}. 

We have modified the Tangled Nature model of evolutionary ecology developed by one of the authors and his collaborators (see e.g. \citet{Jensen_Ann_2010} and references there in). The reason we take our outset in the Tangled Nature framework is that it was demonstrated for the case of evolutionary ecology that this very minimalist approach compares very well with observation. From the simple assumption of mutation prone reproduction of individuals whose reproduction rate depends on the instantaneous configuration in type space, a long list of evolutionary and ecological phenomenology is reproduced.  For example the intermittency of extinction and creation events termed Punctuated Equilibrium by Gouold and Eldredge \cite{Gould77}, realistic species abundance distributions and species area laws. The model can also reproduce relationships between strength of interaction and diversity seen in microbial experiments \cite{DivInt}. This efficiency in obtaining systemic level phenomenology from some basic dynamical assumptions suggests to us that it is worthwhile to explore to what extend a similar framework can be developed for economics.

Since the foundation of the Santa Fe Institute almost three decades a go, researchers have explore the application of complexity science to approach economic problems avoiding the assumption that an economy is a system in equilibrium \cite{arthur1999complexity}. For a complete listing of research in complexity economics see  \citet{kling2011psst} and the references.  Our contribution to this literature is to describe  an evolutionary model of interacting components representing companies' interactions which aggregate  into a model economy that agrees qualitatively with the evolution of the US economy since the great depression, a historical event in time defined by   \citet{kling2011psst} as a restructuring in the world economy that induced a new phase with a new economic structure, which we consider comparable with the beginning of our simulated economy. We find qualitative resemblance in the evolution of the Gross Domestic Product (GDP), the evolution of the number of companies in the economy, and the distribution of company ages and capital. In addition key parameters of the model that can be related to competition laws and the density of company business interactions and leverage, have important effects on the systemic risk of the simulated economy. We suggest that in the future our modeling approach may help to  complement the standard equilibrium theory in macro economics by complexity economics. 

The remainder of this paper is organized as follows: Section 2 presents the model dynamics. In Section 3 we compare the behavior of our model to the US economy from 1929 to 2010. Section 4 presents preliminary predictions from the model regarding systemic risks of the economy with potential policy implications. Section 5 contains a brief discussion and concludes.

\section{The Model Framework} 
\label{Framework}
We use a generalization of the version of the Tangled Nature framework described in Ref. \cite{lair05:tang}. 
A company  $\alpha$ is represented by a string of $L$ traits $T^\alpha=(T_1^\alpha , ..., T_L^\alpha)$ in what we denote the {\em economy space}, and has capital $C^\alpha(t)$ at time $t$.  Here we have used $T_i^\alpha\in\{0,1,2,...,999\}$ with periodic boundary conditions, i.e. there is only one unit distance between 999 and 0. For computational and representational ease in this paper we have set $L=3$. We can, for example, think of the three coordinates as indicative of the intensity in the use of inputs from the agricultural, industry, and services sectors which are embedded in final product of the company. Hence two companies close together in the economy space will produce similar products.   Let us clarify that since the coordinates are periodical, in this interpretation we need to consider the axis of coordinates as a circle and we can, say, think of the companies at the "top of the circle" (e.g. with $T_i^\alpha\approx500$) as  producing a product highly specialized on the given sector. For example, an automobile  company could be high on the industry coordinate circle, at medium hight on the services coordinate circle (e.g. it may have a financing department which lends to buyers, as well as publicity and others services involved), and low on the agriculture coordinate circle (e.g. they may use some wood in their cars). And for instance, say, a computing company will be high on the industry circle but it may be on the opposite side of the circle as the automobile company e.g. the International Space Station can be at the very top of the industry circle with coordinate 500, a car company can be at little lower at coordinate 550, and a computing company can be 450. There are $1000^3$ possible companies in the considered  economy space. 

We use an stochastic sequential update for the capital $C^\alpha(t)$ of company $\alpha$ at period $t+1$ by following the update algorithm described below. One {\em iteration} corresponds to $N(t)$ stochastic updates, where $N(t)$ is the number of companies present at time $t$. The companies affect each other in two different ways: one is doing direct 'business', which we will call interactions, representing anything from buying or selling merchandise between them, to consulting and engaging in financial transactions; the other is competition and applies to companies that are close together in the economy space (i.e. produce similar products). The outcome of the interactions and competition together with the natural resources available in the system determines the probability of gaining or losing capital at the end of the iteration. 

Company $\alpha$ interacts with company $\beta$ with strength $J(\alpha, \beta)$ independent of the corresponding reciprocal interaction strength for $\beta$, $J(\beta, \alpha)$. Only a subset of these $\alpha$-$\beta$ interactions are active (as in a real economy not all companies do businesses together) and the number of potential interactions depends on model parameter $C_{connect}$ (see Appendix A). We compute $J(\alpha, \beta)$ by adding interaction contributions $J_i(T^\alpha_i,T^\beta_i)$ from $L$ different kind of interactions 
\begin{equation}
J(\beta, \alpha)=\frac{1}{L^{1/2}}\sum_{i=1}^LJ_i(T^\alpha_i,T^\beta_i)
\end{equation}
The individual interaction contributions are normally distributed so we use the $\frac{1}{L^{1/2}}$ to ensure the total interaction is similarly distributed. For simplicity we use the same number of individual interaction contributions as the number of coordinates, $L=3$, and one can, for example,  interpret these three individual interactions as exchange of goods, financial transactions and intellectual collaboration respectively. In Appendix A we describe our procedure to obtain an exponentially correlated set of  interactions $J(\alpha,\beta)$.    

 An important property of these interactions is that two companies close together in the economy space have highly correlated interactions with any third company. By construction as the distance between $\alpha$ and $\beta$ increases in the economy space there is an exponentially decay of the correlations  
\begin{equation}
Corr_i(\alpha,\beta) = \left\langle J_i\left(T^\alpha_i,T^\gamma_i\right),  J_i(T^\beta_i,T^\gamma_i)\right\rangle
\end{equation}
for any third company  $\gamma$. This property has a natural interpretation. Say company $\alpha$ interacts with company $\gamma$ through trait $i$ (e.g. exchange of goods), and say company $\beta$ is very close to company $\alpha$ in the economy space (i.e. they produce very similar products), then $T_i^{ \beta \gamma}$ will be very close to $T_i^{\alpha \gamma}$ (see sum (\ref{sum1}) in Appendix A) hence mapping to coordinates close together on the corresponding trait axis, so the $\beta$-$\gamma$ interactions on trait $i$ will be highly correlated to the $\alpha$-$\gamma$ interaction i.e.   $\alpha$ and $\beta$ will benefit (or be disadvantaged) from company $\gamma$ with a similar strength provided that companies $\beta$ and $\gamma$  also do business on trait $i$ (i.e. provided that the corresponding interaction series is active in the trait $T_i^{\beta \gamma }$ so that the interaction $J_i(T^\beta_i,T^\gamma_i)$ is active).  The degree of similarity of companies decays exponentially over a correlation length $\xi$, a parameter of the model set at 300, which characterizes the matrix $J(\beta, \alpha)$. For details see Eq. (\ref{competition}) in Appendix A.

 Based on the property that the similarity of companies' profile decays exponentially with distance in the economy space, we define a {\it direct} competition function $C(\alpha, \beta)$ with a similar exponential decay, in particular with same decay constant $\xi$ as the correlation length of the interaction series (see Appendix A for details).  Companies close together in the economy space are similar due to the previous interaction correlation property, hence they compete strongly, whereas companies more than $\xi$ units away in the economy space have negligible competition as their similarity also becomes negligible. We do acknowledge there may be indirect competition that arises from other economic sectors, that is from a company far a way in the economy space, which will not be capture by this competition function. This common indirect competition are likely to be subject of technological innovation and evolution of the economy, and can be capture by the negative interactions $J(\beta, \alpha)$.
 
The performance of each company at each update is subject to natural resources available in the system, $R(t)$; each company takes one unit of natural resources from $R(t)$ at its creation and when they go bankrupt they return their unit back to the pool of natural resources making the sum $N(t)+ R(t)$ a constant, where $N(t)$ is the number of companies at time $t$. Obviously a more realistic description would be to let the resource bound in a company dependent on the type of company i.e. companies' position in the economy space.  But for simplicity here we assume the resource to be independent of time and type of company. 

Using the total interaction strength, the total competition strength and the natural resources we can now define a weight function $H(\alpha, t)$ that will determine the probability of company $\alpha$ to gain or lose capital:
\begin{equation}
H(\alpha, t)=a_1\frac{\sum_{\beta=1}^{N(t)}J(\alpha, \beta)}{\sum_{\beta=1}^{N(t)}C(\alpha, \beta)}-a_2\sum_{\beta=1}^{N(t)}C(\alpha, \beta)-a_3\frac{N(t)}{R(t)}
\label{weightfct}
\end{equation}
where coefficients $a_i$ are just constants used to ensure the three terms have a similar order of magnitude. The first term represents the effect of interactions with all other companies present in the system at time $t$ (note that many interactions are inactive, i.e. equal to zero). The denominator captures the number of companies in the same economic sector  as company $\alpha$. The rational for this ratio may be better explained by the following example: Say company $\alpha$ produces cars, then all other car companies $\gamma$ are points close to $\alpha$ in the economy space and due to the correlations of the  interaction matrix  the interactions  will have similar total strength $\sum_{\beta=1}^{N(t)}J(\gamma, \beta)$. But usually a given company $\beta$ which is beneficial to company $\alpha$  can not realistically benefit all car companies (with some monopoly exception in the real world) hence to make it more realistic we divide the total interaction by the proxy of the number of companies in the given economic sector.
 The second term in the weight function (\ref{weightfct}) represents the direct competition affecting company $\alpha$, therefore the term is negative capturing the negative impact of  competing companies with the same profile and market aim.   The third term represents the finite amount of resources available to the $N(t)$ companies. The relative strength between these three terms is now used to determine the probability of gaining or losing capital. Namely, we define 
\begin{equation}
P_{gain}=\frac{\exp[H(\alpha, t)]}{1+\exp[H(\alpha, t)]}
\end{equation}
Upon success, company $\alpha$'s capital increases:
\begin{equation}
C^\alpha(t+1)=C^\alpha(t)\left(1+c_g\frac{J^+(\alpha)}{J^{Tot}(\alpha)}\right)
\end{equation}
where $J^+(\alpha)$ is the total strength of positive interactions, $J^{Tot}(\alpha)$ is the total sum of the absolute value of all of $\alpha$'s interactions, and $c_g$ is simply a gain control coefficient to ensure realistic growth.
Similarly, if unsuccessful, company $\alpha$ loses capital according to
\begin{equation}
C^\alpha(t+1)=C^\alpha(t)\left(1-c_l\frac{J^-(\alpha)}{J^{Tot}(\alpha)}\right).
\end{equation}
 
A company whose capital goes below the {\it bankruptcy threshold} is simply removed from the economy and one unit of natural resources is returned to the pool $R(t+1)$. On the other hand if a company's capital goes above the {\it investment threshold} the  company may invests in a new company with constant probability $P_{inv}$. Upon success a new company is created in a random position in the region of size $100\times 100\times 100$ centered around the position of the founder company  in the economy space. The new company starts with $10\%$ of the founder's capital and takes a unit of natural resource from the pool $R(t+1)$. Note that the new company will be well within correlation length $\xi$ (fixed at 300) form the founder company. This means that the new company will have a negative impact on its founder in terms of competition -- although there can be a trade-off in terms of positive interactions. Therefore this feature of the model is to be interpreted as a competition law within a successful market rather than an expansion of the investing company.  For example think of when the US supreme court divided Standard Oil in 1911 because they were dominating the oil market.  We will see that the probability of investment $P_{inv}$ is hence a very important parameter as it controls the distribution of capital and the appearance of oligopolies in the system. We will see in Section 4 details of the impact on the systematic risk of the economy. 

  We let the model run for several thousand iterations and observe the aggregate dynamics that arise from these simple companies interactions.
 
\section{Case Study}
\label{Case}
To assess the usefulness of our simple abstract approach we now make a comparison between dynamical trends in the model and equivalent observables from the the US economy, where data can be readily obtained. 
The model is able to generate aggregate dynamics that are similar to those observed in the US economy. The dynamics of macroeconomic variables are related to key model parameters such as the number of potential business interaction in the economy space ( $C_{connect}$), and the redistribution of capital by the probability of generating new companies ($P_{inv}$). Nonetheless contingency plays a very important role: two independent runs with the same parameters typically exhibit evolutionary paths that differ in details; but this is certainly consistent with experience from the real world.
 
\subsection*{General Macroeconomic Trends: Model v.s. US Economy}
\begin{figure}[h!]
\begin{center}
\includegraphics[scale=0.65]{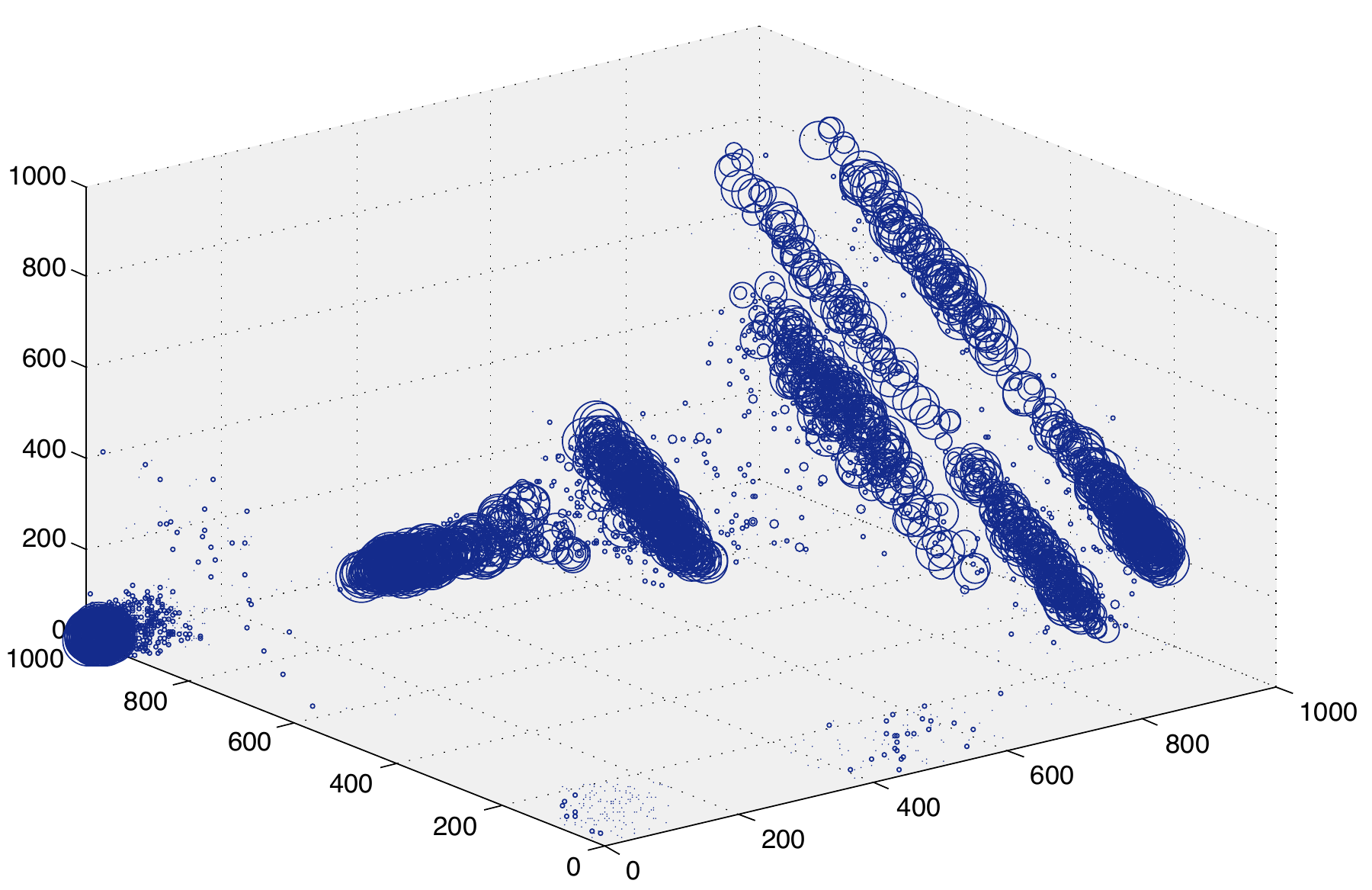}
\caption{ Economy space at t=5000. Companies size proportional to age.}
\label{EcoSp}
\end{center}
\end{figure}

Our starting  configuration consists of 1000 companies placed at random positions in the economy space and with a randomly assigned capital  in the range between 110\% of the Bankruptcy Threshold and 110\% of the Investment Threshold.
After about 1000 iterations this random set of companies, with a given matrix of interactions, self-organizes into well defined structures in the economy space consisting of  tubes and planes (see Figure (\ref{EcoSp})). These structures can be interpreted as the supply chains of given products. Like in real economies who specialize in specific productions, the simulated economy also undergo specialization, which is seen as the formation of coherent structures around specific trait coordinates. Hence, probably not controversial at all, our model indicates that specialization of economies is mainly an emerging property driven by increasing returns and path dependence.

Here we focus in the following macroeconomic variables among other feasible observable variables: GDP (as the proxy for GDP in our model we refer to the sum of the capital of all companies at time $t$), the total number of companies, and cross section distributions of companies' age and capital. We ran 40 independent simulations with the same model parameters but random initial companies, out of which 15 simulations collapsed after a few iterations -- they did not survive the initial chaos. We are interested in the average behavior of the remaining 25 simulations. All graphs referring to the model represent averages of these 25 simulations. The GDP growth rates are computed with a lag of 100 iterations giving realistic values for annual growth rates. We use this comparison to calibrate the time scale of the model against real economical time i.e. think of 100 iterations of the model as representing one-year period in the real world.  We use US data from 1929 to 2010; this period of time for the US economy provides a good comparison since it starts at the beginning of the great depression which brought the economy into chaos and marks a 'new beginning',  a new restructure much like the start of the simulated economy. 

\begin{figure}[h!]
\begin{center}
\includegraphics[scale=0.65]{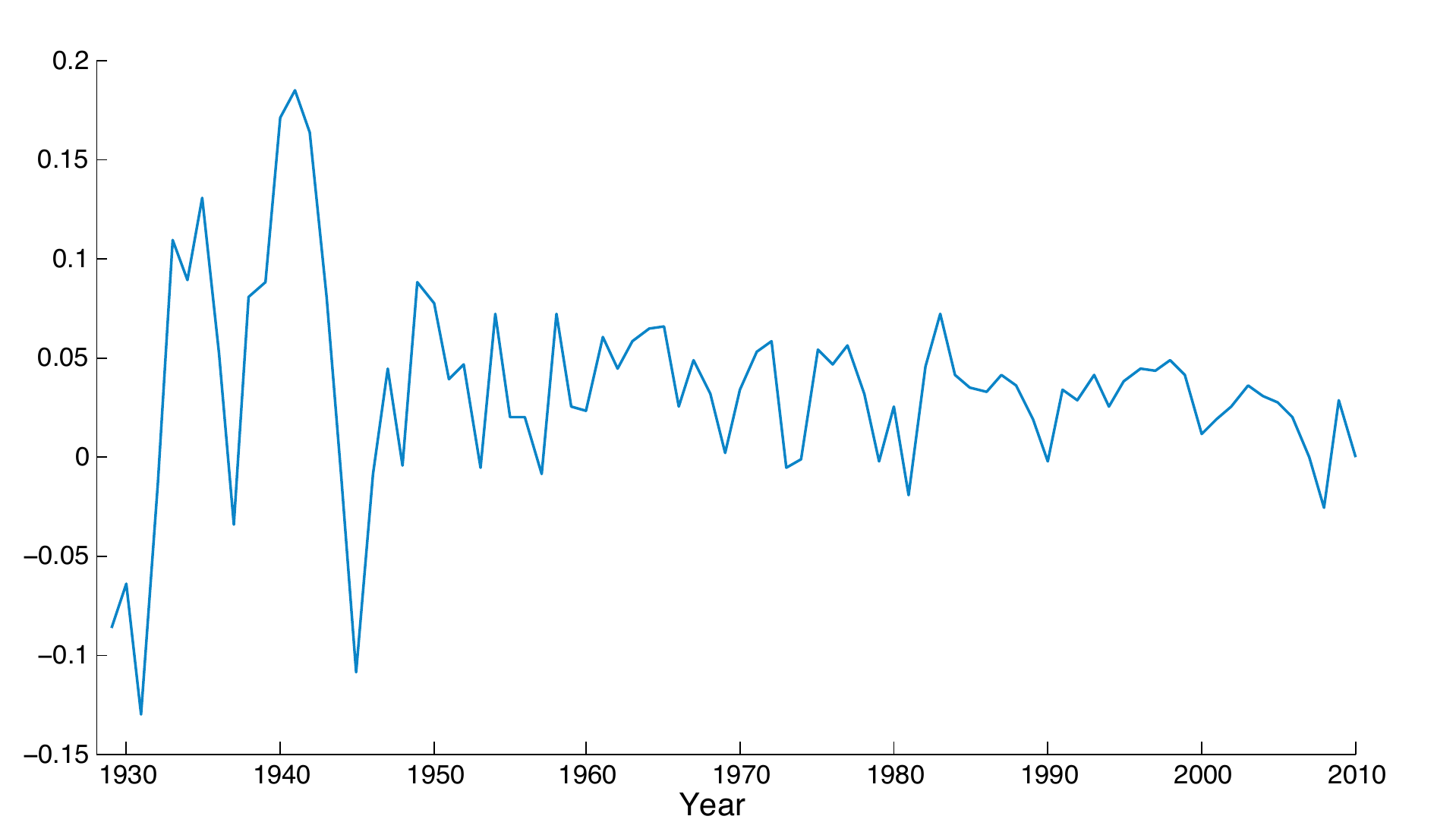}
\caption{   US GDP growth 1929-2010. Corrected for inflation. Source: Bureau of Economic Analysis.    }
\label{GDPgrowth}
\end{center}

\begin{center}
\includegraphics[scale=0.65]{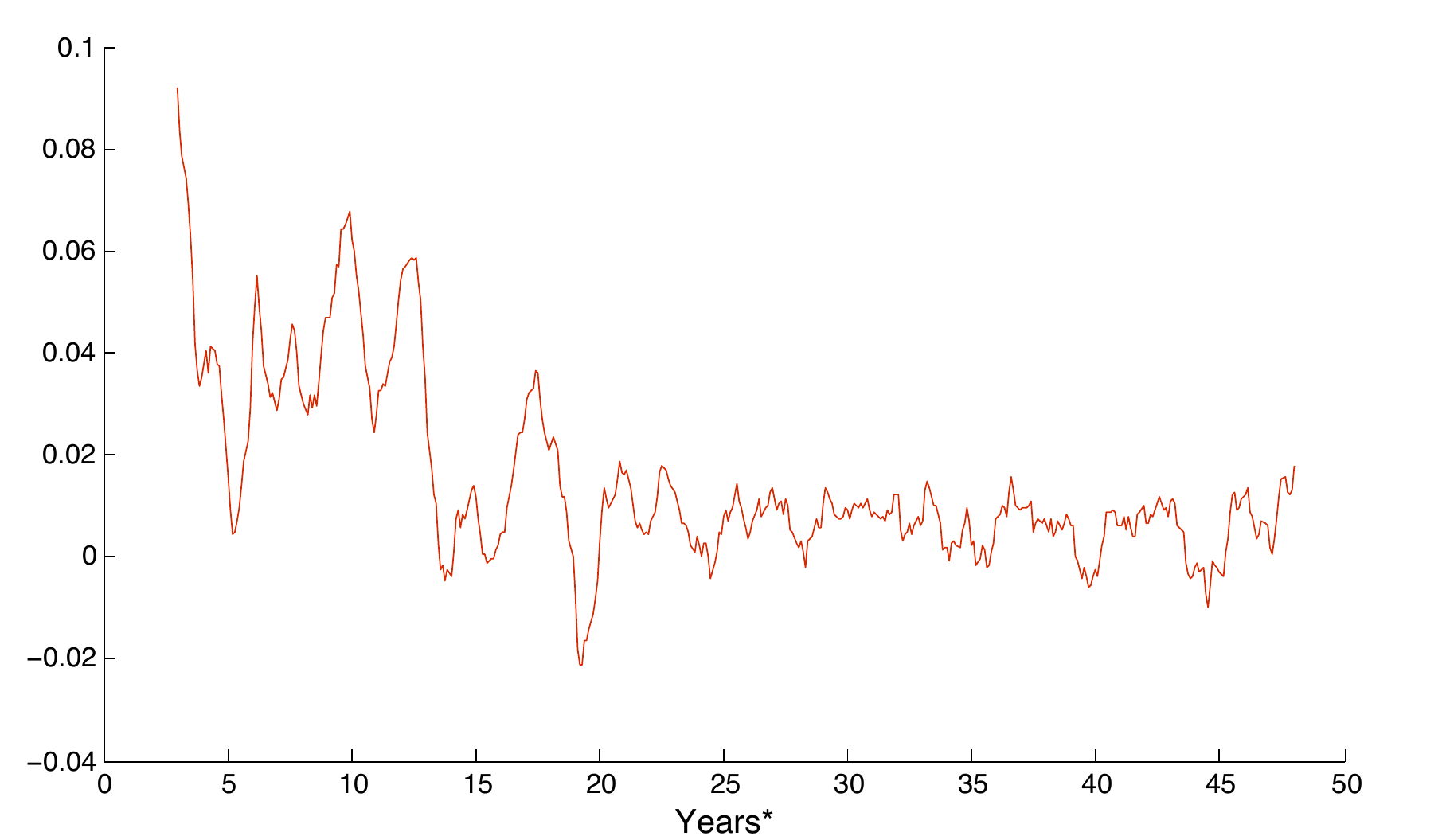}
\caption{Model annual GDP growth. * 1 Year = 100 Iterations. }                          \label{GDPgrowthm}
\end{center}
\end{figure}

\begin{figure}[h!]
\begin{center}
\includegraphics[scale=0.65]{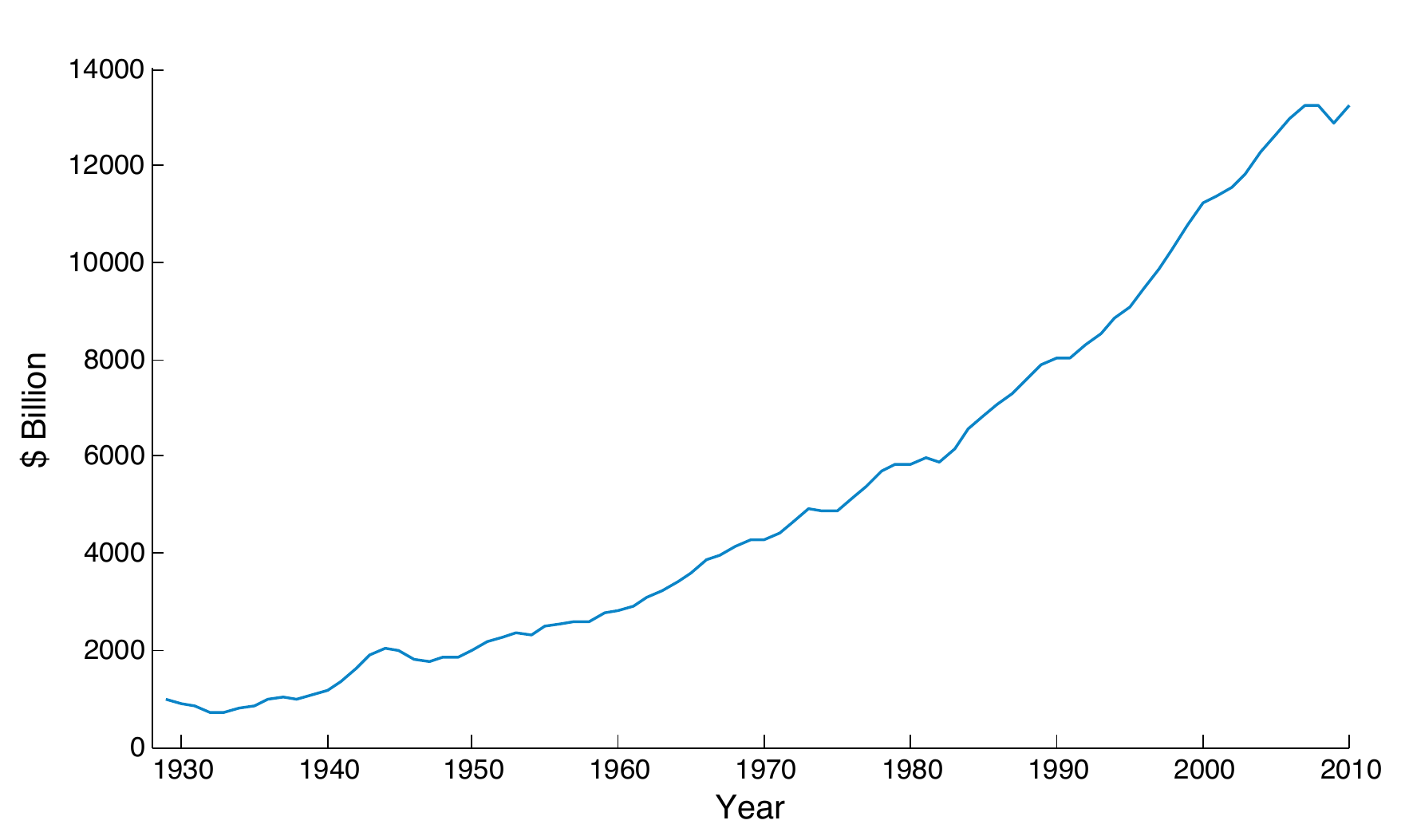}
\caption{US GDP 1929-2010. GDP corrected for inflation. Source: Bureau of Economic Analysis.  }
\label{GDP}
\end{center}
\end{figure}

\begin{figure}[h!]
\begin{center}
\includegraphics[scale=0.65]{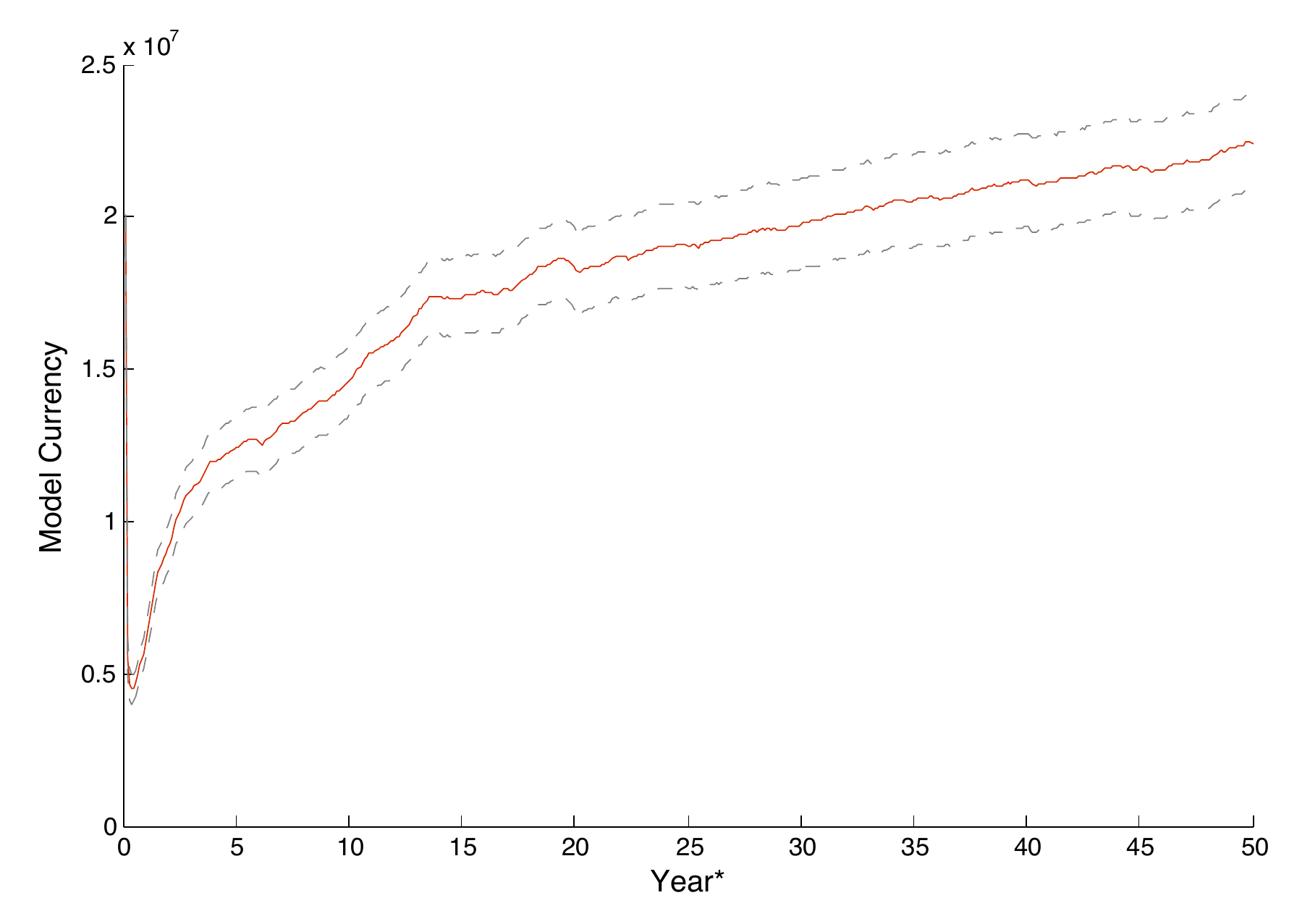}
\caption{Model GDP. Proxy of GDP is sum of all companies' capital. The two dashed curves are the standard deviation divided by 10 away from the averaged given by the solid curve.  The fluctuations are significant.  *We use 1 Year = 100 Iterations, see the text.  }\label{GDPm}
\end{center}
\end{figure}

We find that both the US and the simulated economy have a volatile economic growth at the beginning of the period, with periods of fast GDP growth followed by recession. As both economies mature we observe a damping evolution on their GDP growth rates towards a more modest and stable growth (see Figure (\ref{GDPgrowth}) and (\ref{GDPgrowthm})).  The evolution of the GDP is similar for both economies, both enjoying an overall growth (see Figure  (\ref{GDP}) and (\ref{GDPm})), but an important difference is that the US  GDP follows a convex evolution whereas the simulated GDP tends to be concave. The reason for the concave shape of the simulated GDP comes from the effect of the fixed natural resources limit which restricts the economic growth. 

\begin{figure}[h!]
\begin{center}
\includegraphics[scale=0.65]{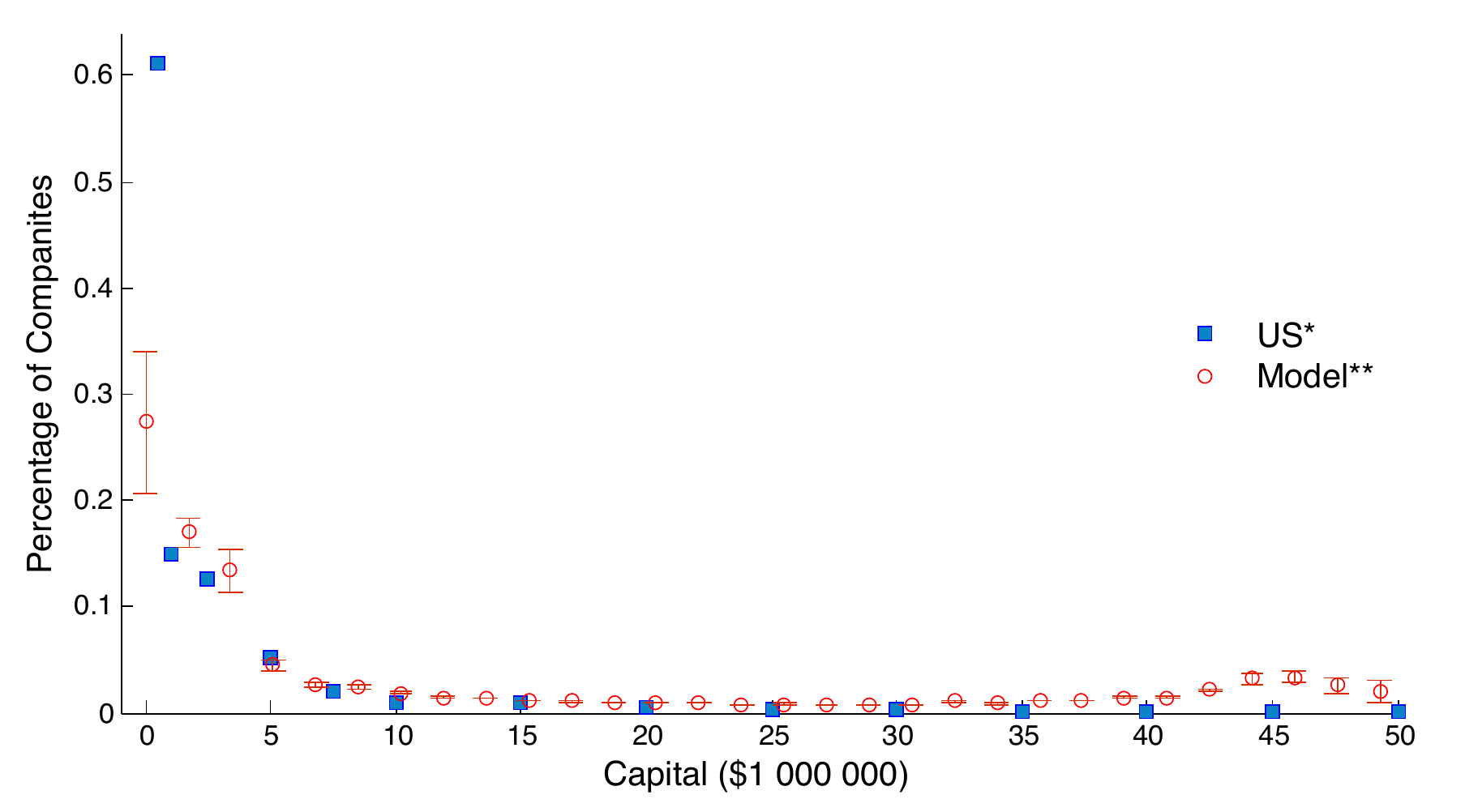}
\caption{Histograms of Companies' Capital. * Source: U.S. Census Bureau 2007. Capital measured by companies' receipts size.                                           
** Model at t = 4900 ;  \$1 = 0.0006 model currency.   }
\label{ModelCapHistCompar}
\end{center}
                
\end{figure}

Another interesting comparison is the distribution of companies size in the model and in the US economy.  Figure (\ref{ModelCapHistCompar})  shows a histogram of US firms in 2007 sorted by receipts size (i.e. expenses) together with model results for the companies capital distribution across the economy at $t=49$ years (note this is already an steady period).\footnote{We chose 2007, and accordingly t=4900 in the model, in order to avoid potential unrepresentative perturbation in the US data by the finantial crisis that started in 2008. Nevertheless there is no qualitative difference when comparing 2009 to t=5000.} Again we notice nice similarities, the model too has many small companies and a few big ones and the proportions are realistic.

\begin{figure}[h!]
\begin{center}
\includegraphics[scale=0.85]{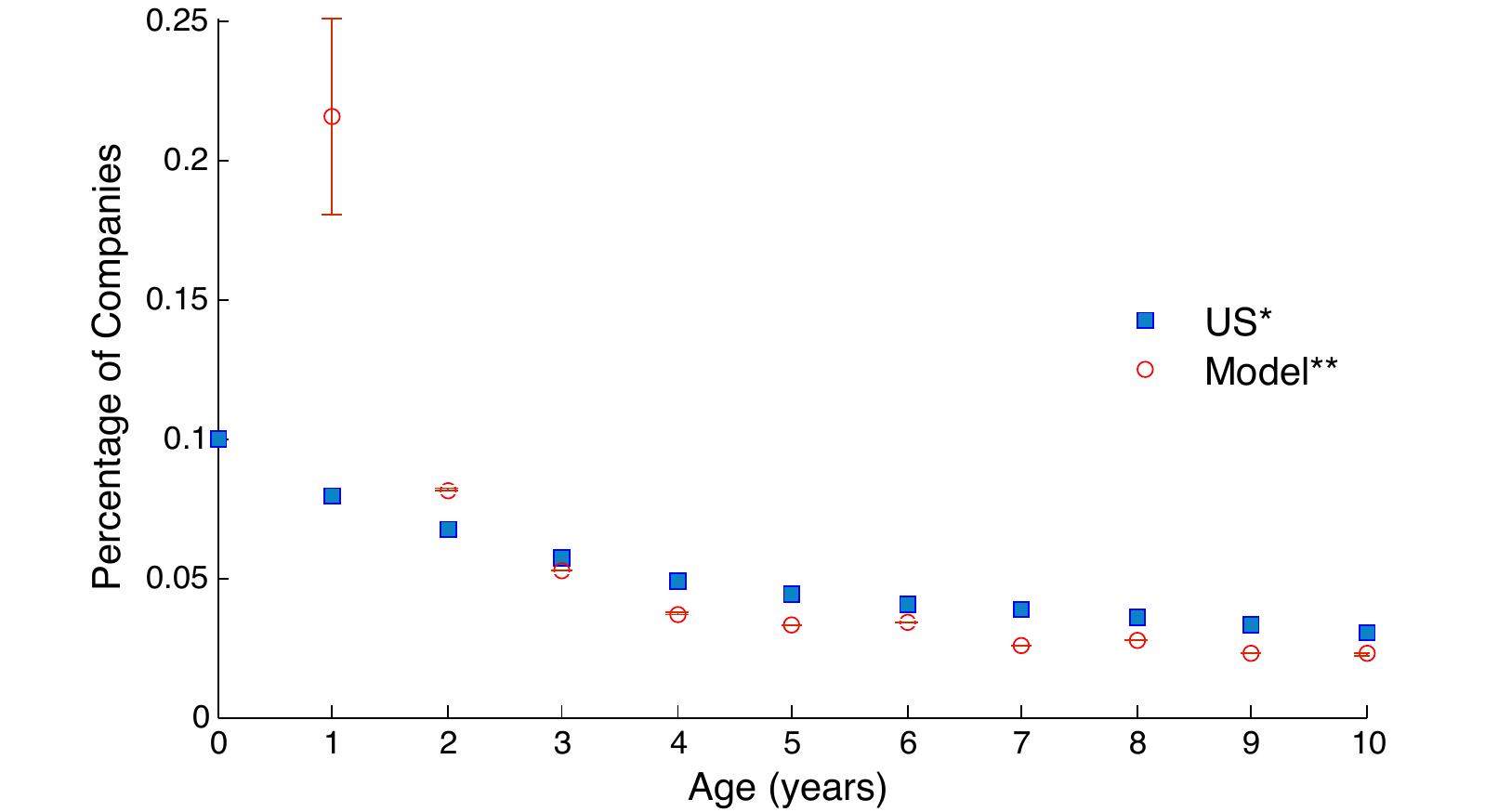}
\caption{Histogram of Companies' Age. *  Source: U.S. Census Bureau 2007; data points for age 6 to 10 are estimates from 5 year average.    
 ** Model at t = 4900 ;  1 Year = 100 Iterations.  }
\label{CompsAge}
\end{center}
\end{figure}

We also compare companies' age distribution across the economy. Figure (\ref{CompsAge}) compares histograms of the US firms by age in 2007 to average age distribution at $t=49$ years.  They both have a decreasing number of companies as the age categories increases. Note that we ignore the first age category  in the model (less than 75 iterations old) because most of these new companies in the model only live for a few iterations. We argue that these companies would not make it to the 'national statistics'. But indeed one of the limitations of our model is that it has unrealistically large creation and destruction rates (some times reaching as much as 200\% per 'year' !). This property can be tuned by manipulating the thresholds for bankruptcies and for creation of new companies through investments. 

\begin{figure}[t!]
\begin{center}
\includegraphics[scale=0.65]{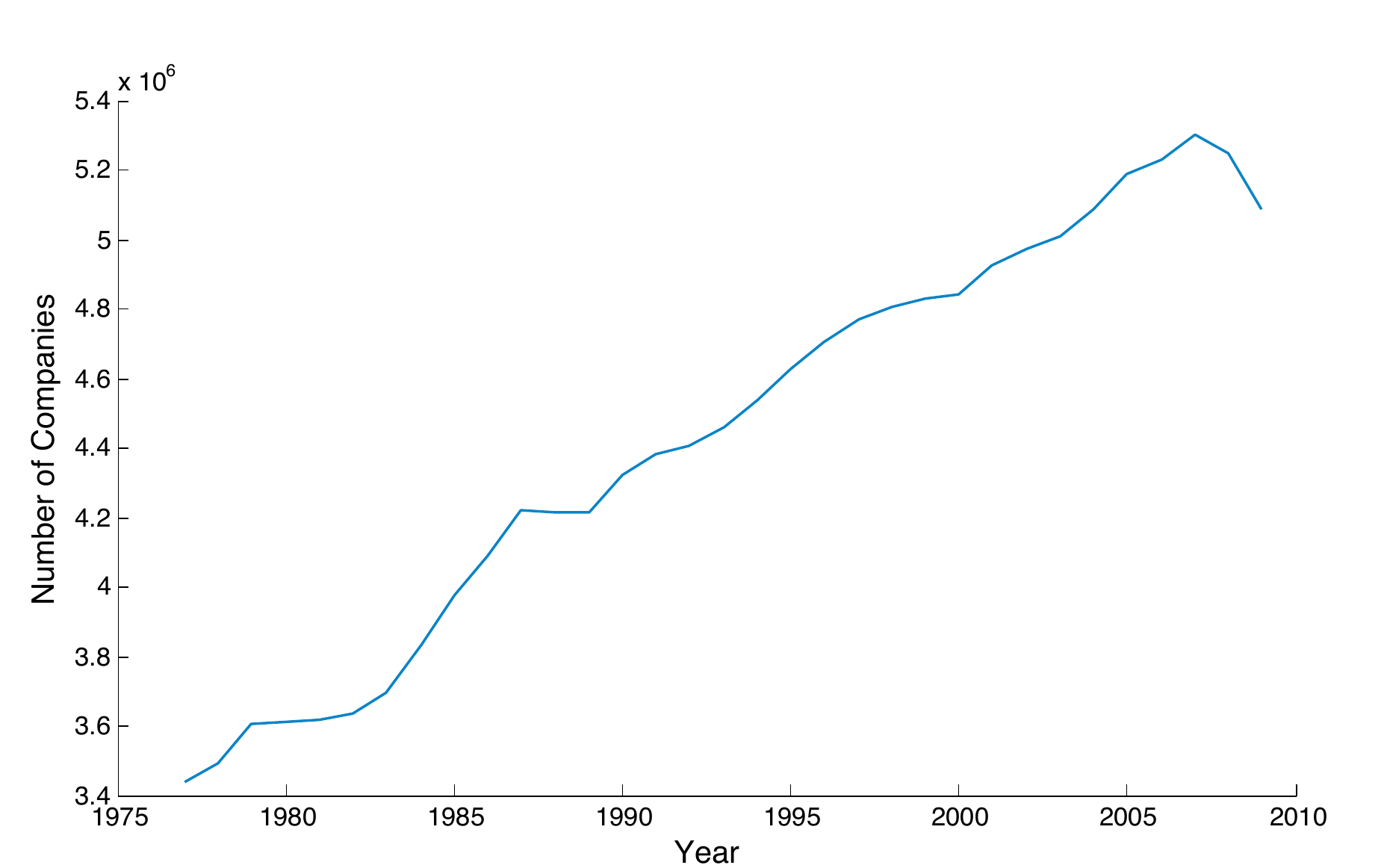}
\caption{Number of Companies in US 1977-2009. Source: U.S. Census Bureau.}
\label{NumCopUS}
\end{center}
\end{figure}

\begin{figure}[h!]
\begin{center}
\includegraphics[scale=0.64]{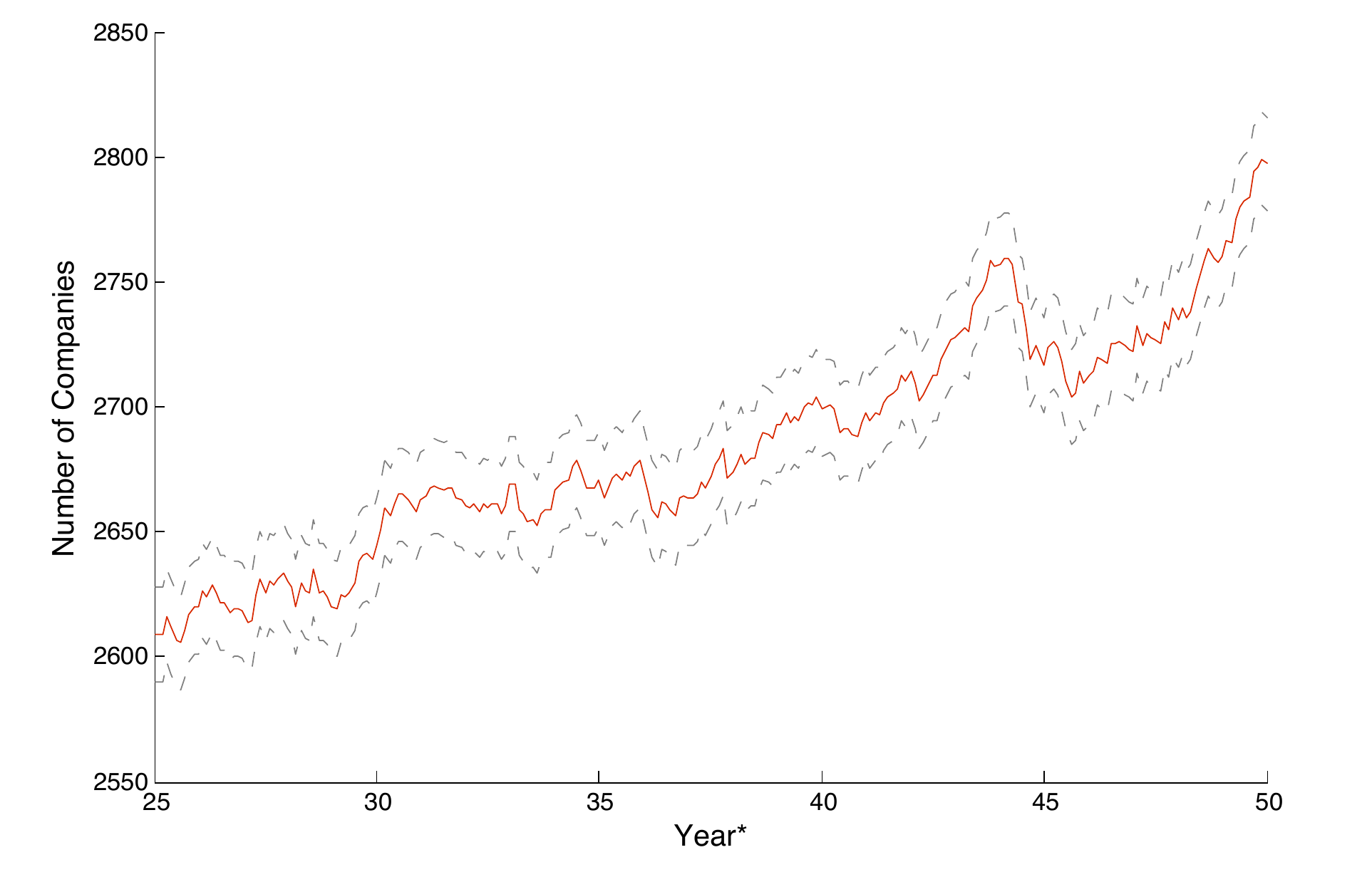}
\caption{Model evolution of number of companies. The two dashed curves are the standard deviation divided by 100 away from the averaged given by the solid curve.  The fluctuations are significant. 
* We use 1 year = 100 iterations. The x-axis represents years since the start of the simulation.}
\label{ModelNumComps}
\end{center}
\end{figure}

Finally the evolution of the number of US firms from 1977 to 2010 and the evolution of the number of companies in the model are shown in Figure (\ref{NumCopUS}) and (\ref{ModelNumComps}). The data of the US starts in 1977 hence for consistence with previous interpretations we compare with the last 2500 iterations of the model where we have a more stable growth similar to the one observed in the US data. The model has a similar tendency although it has more variance due to the problem of excessively large companies destruction and creation rates previously mentioned.

\section{Systemic risk of the model}

We would like to explore the potential use of this model. To this end we analyze how the model dynamics change when we modify two key parameters: The first one is the probability of investment $P_{inv}$, which is related with the distribution of capital in the economy, much like a fair competition law; the second parameter is the connectivity threshold of the interaction series, $C_{connect}$, which is related to the number of potential interactions between companies in the economy space and may be related to a confident/reheated economy for high connectivity, and vice-versa, and also related to companies leverage regulations. We run several simulations for the following values of $P_{inv}=\{0.1; 0. 3; 0.5\}$ and with $C_{connect}=\{-10; 0; 10\}$ (note that in this part of the analysis we did not use averages of multiple simulations due to time constraints, but the patterns we will discuss are very consistent within the simulations we ran for each parameter combination, that is at least three simulations per parameter combination).\footnote{Recall from the model dynamics in Section 1 that lower connectivity threshold gives rise to high number of potential interactions in the system. Hence we will refer to scenarios with $C_{connect}=-10$ as high connectivity,  $C_{connect}=0$ as medium connectivity and $C_{connect}=10$ as low connectivity.} To quantify stability and risk of the system we consider the mean and the volatility (i.e. the variance) of the growth rates of macroeconomic variables: GDP, number of companies, average capital per company, and companies creation/destruction. Due to space constraints   we will not  go into details of the results, but only discuss the volatility of GDP to describe the general behavior of the model. 
 
\begin{figure}[h!]
\begin{center}
\includegraphics[scale=0.65]{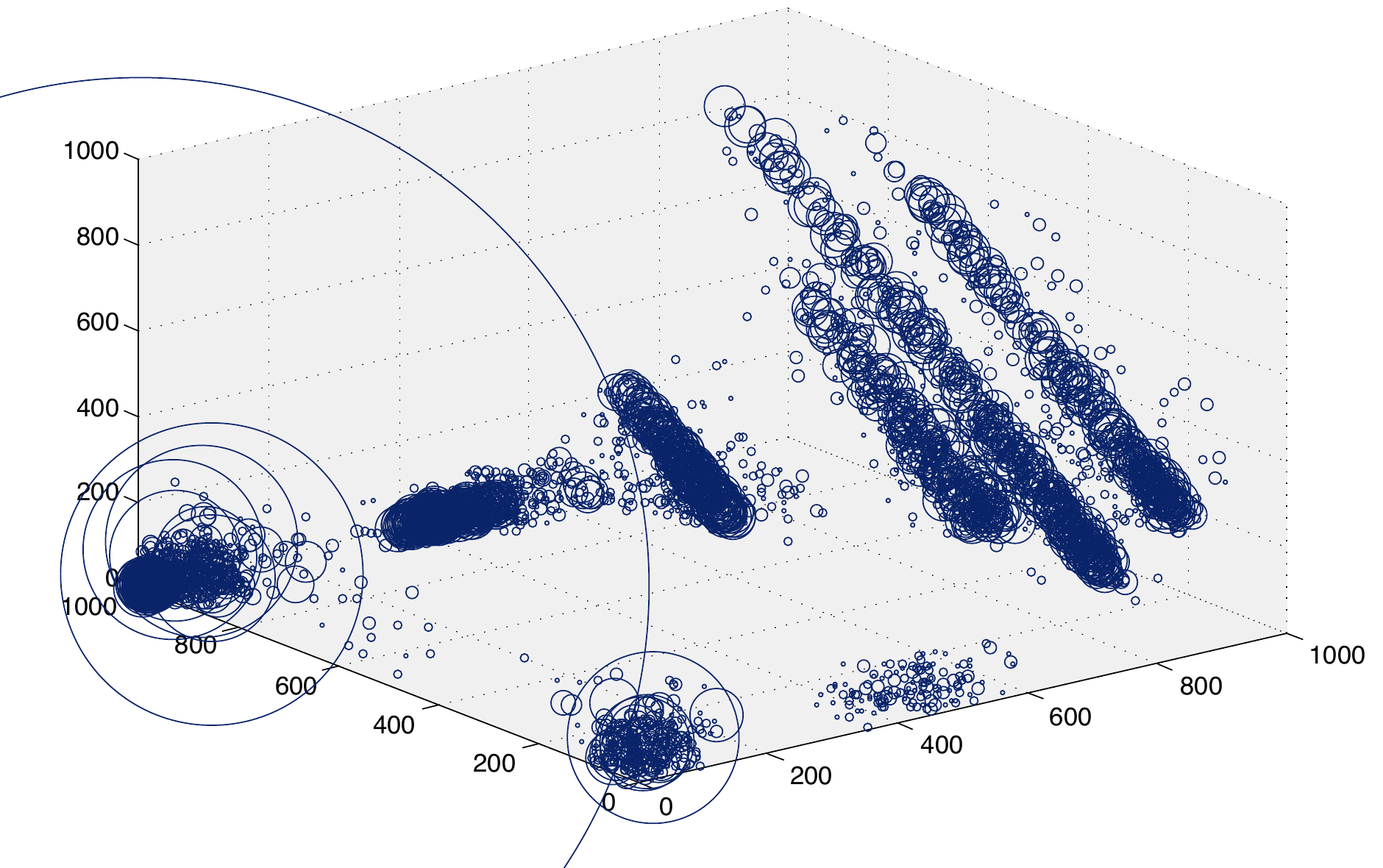}
\caption{Economy space at t=5000. Companies size proportional to capital. $P_{inv}=0.3$; $C_{connect}=0$.}
\label{Oligo}
\end{center}
\end{figure}

We found that an economy with loose fair competition regulations (i.e. low probability of investment $P_{inv}$) increases the average GDP growth but also increases the volatility of this variable inducing strong recessions in the simulated economy. Whereas high probability of investment generates a more modest and stable GDP growth. Similarly a reheated economy with loose leverage regulations (i.e. highly connected economy) tends to have a strong and volatile GDP growth, and low connectivity induces modest but stable growth. As we test different combinations of these two parameters we find what we would expect: deregulation in competition laws in a reheated economy i.e. low probability of investment in a highly connected economy, is the most unstable combination; and strong competition regulation together with leverage regulation i.e. high probability of investment in an economy with low connectivity, is the most stable and modest growth. Further more, the unstable parameter combinations tend to develop oligopolies in the economy, an small number of companies that completely dominate given economic sectors (see Figure (\ref{Oligo}) ). This explains partly why these scenarios are unstable, as the relevant economies are highly dependent on few companies, whereas in the economies which favor a fair competition, that is the stable scenarios, the economic growth relays on many companies so that the collapse of a single company is almost unnoticed. 

This results are not surprising. With low probability for investment companies can grow much bigger, and some eventually become oligopolies or even monopolies with no regulation to control them. This is further facilitated in a highly connected economy were key positions will be at the center of a large web of business, hence companies at advantageous locations are likely to become very successful, although still liable to chance, and the fate of the economy ends up in the hands of few companies. High conectivity  also makes the economy more interdependent and the bad fate of few companies can expand trough the economy with easy generating economic recessions.   Hence arguably we could identify the combination of unstable parameters as corresponding to a deregulated economical policy. Both in terms of lacking fair competition regulations (low probability of investment), and allowing high leverage for companies and deregulated banking system in terms of high number of loans and debts (high connectivity).\footnote{Note that our result concerning the effect of connectivity on the systemic risk of the economy is consistent with the finding of  \citet{Haldane2011}. } Our model is still to simplistic to provide policy advice but is very encouraging that it agrees with well known risk of deregulated economies and furthermore it captures very well the tradeoff between higher average growth and higher systemic risk from deregulation. 

In the future such models could be tuned with real data from an economy taking in account major agents such as big banks and corporations in order to start the simulation with the present state of the economy. From there we could run several simulations and test several policy strategies, and quantify the performance of potential policies in a way that classic economics, in particular macro economics and the general equilibrium theory, cannot.

 \section*{Discussion}
The simple model developed here is able to generate aggregate dynamics for key economic variables that are generally consistent with empirical evidence.   There are nonetheless several limitations and opportunities for improvement. One issue is the effect of the natural resources limit which tends to give a concave shape to the evolution of the GDP and of the number of companies in the model, in contrast with the convex shape shown by the US GDP. We notice that the natural resource limitation of the US is less of a barrier since on top of their national natural resources they can also import from other countries. We could design extensions of the model modifying the role of natural resources in the dynamics to try to capture better this fact  and to allow for technological innovation.
 
As in the real world contingency plays also an important role in the model: two independent runs with the same parameters typically exhibit evolutionary paths that differ in details. This emphasizes the importance of having adaptive policies that are able to introduce real time corrections when a given economic system starts to deviate from a stable path. The other important observation in the model relates to patterns of specialization. Stable economies are not good at producing ÒallÓ goods. Companies end-up operating in given regions of the economy space suggesting that the comparative advantages of an economy are not an exogenous but an emerging property in an economy. Finally, the model predicts the emergence of oligopolies and monopolies as a natural feature of deregulated economies. This also calls for policy interventions that regulate the level of competition, such as anti-trust law. 

We believe the importance of our model is that it highlights how the macro behavior, i.e. the growth or decline of companies, may originate in the {\em entangled} nature of the interaction between companies. We have deliberately kept the internal structure of the individual companies very simple in order to focus on the fact that a given company always finds itself embedded in an ever changing environment composed of all the other co-existing companies. This fact suggests that the notion that a company may possess an intrinsic robustness or fitness is to an extend an illusion.

In the future we plan to study the dynamics presented here on economical networks extracted from real data sets of the network of transactions between companies. With increasing computer power and more sophisticated complexity models these approaches may become key in the future of  macroeconomic policy design.

\section*{Figures}
 
\FloatBarrier
\appendix
\section{}
\begin{figure}[h!]
\begin{center}
\includegraphics[scale=0.73]{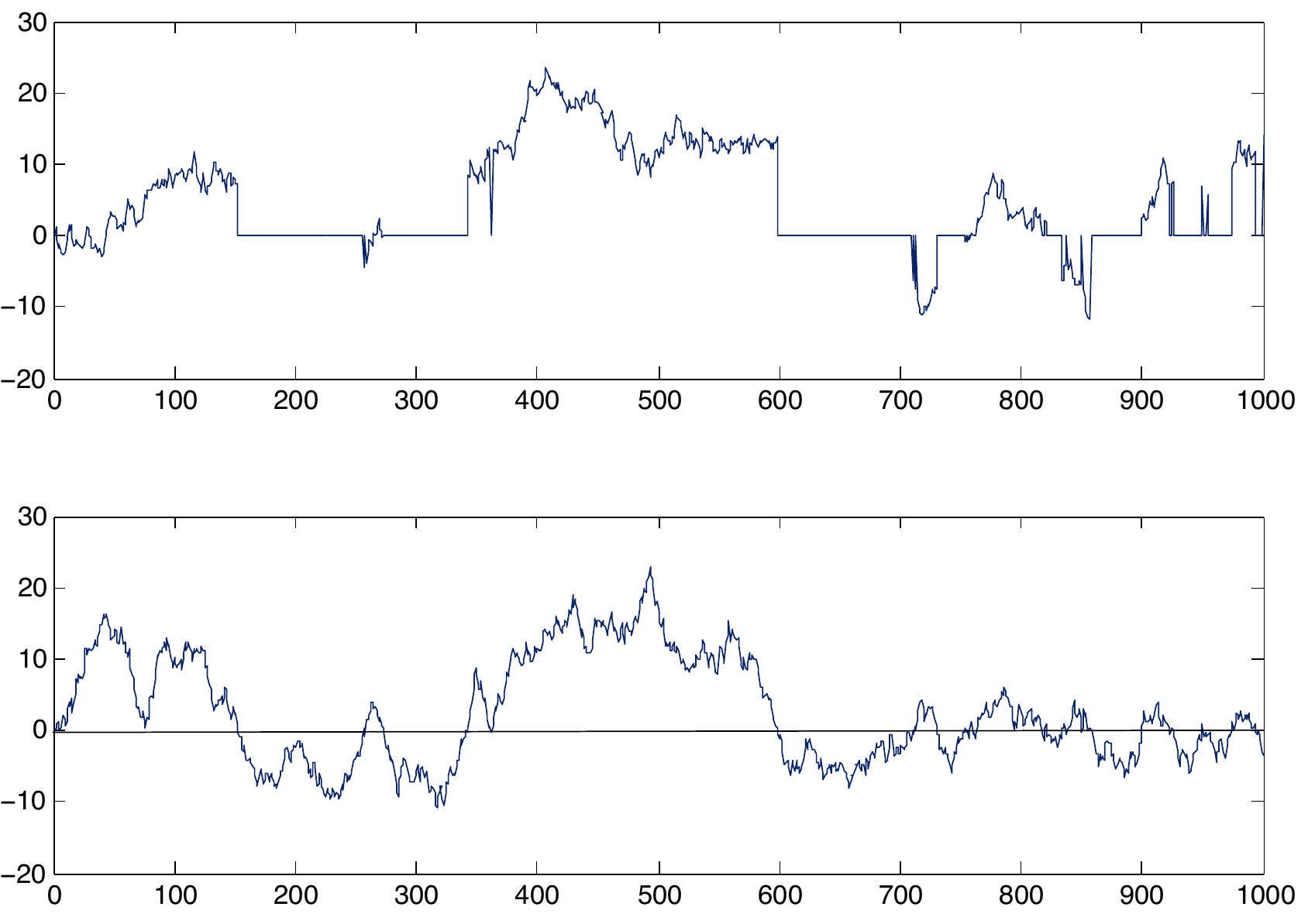}
\caption{Interaction series (top) and corresponding switch series (bottom) with $C_{connect}=0$.}
\label{InterS}
\end{center}
\end{figure}

To compute the individual interactions $J_i(T^\alpha_i,T^\beta_i)$ we use a method developed by \citet{lair05:tang}. We use $L$ independent autoregressive time series, each one associated to an individual interaction: 
\begin{equation}
x_{n+1}=\rho x_n + \phi
\label{ts1}
\end{equation}
where $\phi$ is a normal distributed random shock and $\rho$ is a constant that determines the correlation length of the series. 
We then map the following sum to the corresponding time series
\begin{equation}
T_i^{\alpha \beta}= \sum^L_{j=1}b_{ij}T^\alpha_j + \sum^L_{k=1} c_{ik}T^\beta_k  mod(1000)
\label{sum1}
\end{equation}
where $b_{ij}, c_{ij} \in \{-1,1\}$ are independent randomly assigned and kept fixed trough out the dynamics.
The individual interaction $i$ will be the value of the time series corresponding to the index $T_i^{\alpha \beta}$ (i.e.  $J_i(T^\alpha_i,T^\beta_i)=x_{T_i^{\alpha \beta}}$ belonging to the i$^{th}$ series). We will refer to these three time series as the set of {\it interaction series}. To create the subset of inactive interactions, for each {\it interaction series}, we create a parallel independent {\it 'switch'} time series and whenever this second time series goes below a given {\it connectivity threshold}, $C_{connect}$, the corresponding values of the interaction time series are set to zero (see Figure (\ref{InterS})). This connectivity threshold is important as it will influences the number of potential interactions in the economy (i.e the number of connected points in the economy space).

The motivation to use sum (\ref{sum1}) and the auto-regressive time series for the interactions comes from the properties of the autocorrelation of the time series: From equation (\ref{ts1}) using induction we can write
\begin{eqnarray*}
x_{n}&=&\rho x_{n-1} + \phi_{n}\\
x_{n}&=&\rho (\rho x_{n-2} + \phi_{n-1}) + \phi_n\\
x_{n}&=&\rho^2x_{n-2} +\rho \phi_{n-1} + \phi_n\\
...&&\\
x_{n}&=&\sum_{k=0}^\infty \rho^k \phi_{n-k} \\
\end{eqnarray*}
assuming $x_N=0$ and taking $N \rightarrow -\infty $. Now $\{\phi_i\}$ are IID normal N(0,1), therefore clearly we have $E(x_n)=0$, also $cov(\phi_i, \phi_j)=0$ for $i\neq j$ and $E( \phi_{n-k}^2)=var(\phi_{n-k})=1$ hence we have:
\begin{eqnarray*}
var(x_n)=E(x_n^2)=\sum_{k=0}^\infty \rho^{2k}E( \phi_{n-k}^2)= \frac{1}{1-\rho^2}
\end{eqnarray*}
provided $\rho<1$ (which we ensure in the model). Also $\rho<1$ implies that the process is stationary hence the auto-covariance and autocorrelation between $x_n$ and $x_m$ depend only on the separation length $\tau=|n-m|$. Define $s_\tau=cov(x_n,x_{n-\tau})$, we have: 
\begin{eqnarray*}
x_{n}&=&\rho x_{n-1} + \phi_{n}\\
x_{n}x_{n-\tau}&=&\rho x_{n-1}x_{n-\tau} + \phi_{n}x_{n-\tau}\\
\end{eqnarray*}
taking expectation
\begin{eqnarray*}
s_\tau&=&E(x_nx_{n-\tau})=\rho E(x_{n-1}x_{n-\tau})\\
&=&\rho s_{\tau-1}\\
...&&\\ &=&\rho^\tau s_0\end{eqnarray*}
using induction and the fact that $x_{n-\tau}$ depends only on $\phi_{n-\tau}$ and earlier shocks therefore $x_{n-\tau}$ is uncorrelated with latter $\phi_i$'s (i.e. $E(\phi_nx_{n-\tau})=0$ for $\tau \geq 1$) . 
Finally the most important property of the AR(1) process for the purpose of this model is the autocorrelation sequence which takes the following form
$$Corr(x_n,x_{n-\tau})=\frac{s_\tau}{s_0}=\rho^\tau$$
Hence the autocorrelation of the time series has an exponential decay
\begin{eqnarray}
Corr(x_n,x_m)&=&\rho^{|n-m|}\\ \nonumber
&=&\exp\left[-\frac{|n-m|}{\xi}\right]
\end{eqnarray}
For the purpose of this model we choose the correlation length $\xi$  and we set
$$
\rho=\exp[-1/\xi]
$$

Since the interaction behavior will be considerably correlated for companies closer than $\xi$ units away on each coordinate, we consider these companies as producing similar products, hence the correlation length represents the {\it competition length} in the economy space i.e. the distance between companies at which they should have a non-negligible competition between them. But to define the competition strength, instead of using correlations for each interaction series, we define a single function representing an exponentially decaying competition strength that approximates a similar two point correlation function determined by the separation of the two companies in the economy space:
\begin{equation}
C(\alpha, \beta)=\exp\left[\frac{(-1/L)\sum^L_{i=1}\Delta T^{\alpha \beta}_i}{\xi}\right]
\label{competition}
\end{equation}
where
$$
\Delta T_i^{\alpha \beta}=|b_{i1}(T_1^{\alpha}-T_1^{\beta})+b_{i2}(T_2^{\alpha}-T_2^{\beta})+...+b_{iL}(T_L^{\alpha}-T_L^{\beta})|
$$
 and $b_{ij}$ are the same we use in sum (\ref{sum1}). Self-interactions and self-competition are dismissed.












\end{document}